\documentstyle[12pt]{report}
\begin{document}
\baselineskip=0.8cm
\ \ \

\begin{center}
{\bf STELLAR POPULATION GRADIENTS IN BRIGHT CLUSTER GALAXIES AT z=0.2}
\end{center}

\begin{center}
T. J. Davidge, Gemini Canada Project Office,

Dominion Astrophysical Observatory, 5071 West Saanich Road,

Victoria, B. C., CANADA V8X 4M6$^a$

and

Department of Geophysics \& Astronomy, University of British Columbia,

Vancouver, B. C., CANADA V6T 1Z4

email: davidge@dao.nrc.ca
\end{center}

\begin{center}
M. Grinder, Co-op Program, Department of Physics \& Astronomy,

University of Victoria, Victoria, B. C. CANADA V8W 3P7
\end{center}

\vspace{1.0cm}
\noindent{$^a$ Postal Address}

\begin{center}
{\it To appear in The Astronomical Journal}
\end{center}

\pagebreak[4]
\begin{center}
ABSTRACT
\end{center}

	Slit spectra, covering the rest frame near-ultraviolet and blue
wavelength regions, are combined with moderately deep
$g$ and $R$ images to investigate radial population gradients
in the brightest components of six z=0.2 galaxy clusters selected according to
x-ray brightness. We conclude that the brightest members of the EMSS0440+02,
EMSS0906+11, and EMSS1231+15 clusters have {\it global} stellar contents
similar to nearby elliptical galaxies, although all three galaxies studied in
EMSS0440 contain a central blue component, suggesting that a young or
intermediate-age population is also present. Our ability to investigate the
stellar content of the brightest cluster galaxies (BCG's) in the
clusters EMSS0839+29, EMSS1455+22, and Abell2390 is complicated by ionized
gas, which is detected over large portions of all three systems.
Radial color variations are detected in the majority of galaxies and,
for those systems not showing extended [OII] emission,
there is a tendency for the 4000\AA\ break to weaken with increasing radius,
such that $\Delta$D4000/$\Delta$log(r), where D4000 measures the
strength of the 4000\AA\ break in magnitudes, falls near the upper range of
what is seen among nearby ellipticals. However, there is no correlation between
local D4000 values and $(g-r)$, suggesting that these quantities are sensitive
to different parameters. There is a slight tendency
for the steepest spectroscopic gradients to occur in the BCG's. We
estimate the metallicity sensitivity of the 4000\AA\ break
using a combination of observational and theoretical data, and
find that $\Delta$D4000/$\Delta$[Fe/H] $\sim 0.4 - 0.8$ in the super metal-rich
regime. The metallicity gradients derived for the BCG's from this calibration
are relatively steep, and are in qualitative agreement with predictions
for galaxies which experience a single monolithic dissipative collapse.
Line gradients measured with respect to $g$ surface brightness,
as opposed to radius, show excellent galaxy-to-galaxy consistency,
and are in good agreement with what has been measured in Coma cluster
galaxies, suggesting that common physical processes controlled the early
phases of star formation and chemical enrichment in
these systems. The presence of significant stellar population
gradients in the BCG's suggests that the material from which they formed
experienced dissipation during the collapse phases, and it seems likely that
these galaxies could not have formed solely from the merging of gas-poor
galaxies.

\pagebreak[4]
\begin{center}
1. INTRODUCTION
\end{center}

	It has long been realised that brightest cluster galaxies (BCG's) are
important cosmological probes. For example, it is now well established that
BCG's at low redshift are standard candles (e.g. Sandage \&
Hardy 1973; Hoessel, Gunn, \& Thuan 1980), and hence are useful for studies of
large-scale deviations from the Hubble flow (e.g. Lauer \&
Postman 1992). Moreover, as the largest galaxies, BCG's provide a natural means
of studying the relationship between structures on galactic and cluster scales.
In fact, there is a clear tendency for the axes of BCG's to align
with those of their host clusters (e.g. Sastry 1968; Carter \& Metcalfe 1981;
Bingelli 1982; Struble 1987; Porter, Schneider, \& Hoessel 1991), and possibly
even point to other clusters (e.g. Rhee, van Haarlem, \& Katgert 1992).

	Given the importance of BCG's as cosmological probes, it is crucial to
develop an understanding of their evolution. There is no reason {\it a priori}
to expect that BCG's should follow evolutionary paths similar to those of
nearby classical elliptical galaxies, as BCG's occur in dense, kinematically
hot environments. Nevertheless, the structural properties (e.g. Ryden,
Lauer, \& Postman 1993; Oegerle \& Hoessel 1991, but see also Porter et al.
1991) and integrated stellar contents (e.g. Schneider, Gunn, \& Hoessel 1983a)
of BCG's are ostentsibly similar to those of classical elliptical galaxies.

	A fundamental issue which has yet to be resolved is the time scale of
BCG formation, especially when measured with respect to lower mass cluster
members. At one extreme it has been suggested that BCG's formed over an
extended period of time as smaller systems merged near cluster centers;
however, it has also been argued that the main body of BCG's could only have
formed during early epochs. Some studies (e.g. Bhavser 1989) conclude that
more than one process may be responsible for the formation of BCG's.

	The evidence for on-going mergers is substantial, and will only
briefly be reviewed here. Simple merger simulations can successfully reproduce
many properties of BCG's, such as their extended sizes,
and their use as standard candles (e.g. Hausman
\& Ostriker 1978). Detailed hydrodynamical simulations of cluster
evolution, which reproduce global cluster properties such
as total x-ray flux, form BCG's from mergers which
terminate only near the present epoch (e.g. Katz \& White 1993).
There is also photometric (e.g. Schneider, Gunn, \& Hoessel 1983b) and
structural (e.g. Ryden et al. 1993) evidence from studies of the
second and third-ranked cluster ellipticals that the evolution of BCG's occurs
at the expense of other cluster members, as might be anticipated if mergers
have occured. Finally, roughly one sixth of nearby
BCG's show low surface-brightness companions which may be short-lived artifacts
of tidal interactions (e.g. Porter et al. 1991).

	However, despite being able to predict many of the observed
properties of BCG's and their companions, models which assume extensive recent
merger activity are not free of difficulties. Merritt (1985) concludes that the
incidence of tidal capture is rare in present-day clusters, unless (1) the
galaxies retain substantial dark matter halos, which seems unlikely due to
tidal truncation, and/or (2) the cluster velocity dispersion is low. Merritt
speculates that the main body of BCG's formed during early epochs, when
physical conditions in the cluster were very different from those prevalent
today. It should be emphasized that recent episodes of merging or tidal
stripping are not completely precluded; rather, these events are rare, and will
not contribute substantial mass to BCG's at the current epoch. Observational
support that BCG's do not experience a significant number of mergers at current
epochs comes from dynamical studies of their nearest neighbors, which indicate
that many of these objects are not bound to the central galaxy in circular
orbits (e.g. Tonry 1985; Blakeslee \& Tonry 1992). Finally, McLaughlin, Harris,
\& Hanes (1993) and Lee \& Geisler (1993) have investigated the properties of
globular clusters projected against the inner and outer regions of M87, a
galaxy
with characteristics reminiscent of cD systems, and find that many properties
of the cluster system do not change with radius. This suggests that
the globular cluster system, and by inference the main body of
the galaxy, formed in a homogeneous fashion, presumably during early epochs.

	In this paper we investigate BCG evolution by studying a sample of
objects at intermediate redshift. In particular, broad-band imaging and slit
spectroscopy of six z=0.2 BCG's and bright companions in clusters selected
according to x-ray luminosity are used to search
for stellar population gradients out to $\sim 5$ arcsec ($\sim 7$ kpc at
z=0.2 if H$_0 \sim 75$km/sec/Mpc and q$_0 \sim 0$) from the galaxy centers.
The study of population gradients is of interest to BCG evolution as it
provides information concerning the collapse history of these objects. Indeed,
some merger models predict that stellar content will not change
with radius, while the detection of spectroscopic and/or color gradients
similar in strength to those seen in nearby
ellipticals would be suggestive of common formation mechanisms.
The observations and the reduction of the data are
discussed in Section 2, while the measurements of line indices and colors are
discussed in Sections 3 and 4, respectively. The characteristics of the
photometric and spectroscopic gradients
are compared in Section 5. A discussion and summary of
the results follows in Sections 6 and 7, respectively.

\begin{center}
2. OBSERVATIONS AND REDUCTIONS
\end{center}

	The data were recorded at the f/8 focus of the 3.6 metre
Canada-France-Hawaii Telescope (CFHT) using the MOS arm of the
MOS/SIS spectrograph (Le F\`{e}vre et al. 1993; Morbey 1992). The detector was
Loral3, an uncoated thick CCD with 15$\mu$m pixels in a 2048 x 2048 format,
which has a spatial scale of 0.31 arcsec/pixel at the MOS focus. The O300 grism
(300 grooves/mm; $\lambda_{blaze} \sim 5800$\AA), which produces a dispersion
of 3.45 \AA/pixel, was used as the dispersing element; the
spectral resolution, measured from the width of arc lines, is
15.5\AA, consistent with the projected widths of the 1.5 arcsec slits.

	The data discussed in this paper were obtained as part of the CNOC
survey to investigate the dynamics of an x-ray selected sample of galaxy
clusters (Carlberg et al. 1994). The main observational goal of the CNOC
program was to obtain a large database of galaxy redshifts, and this defined
key observational parameters such as wavelength coverage and integration time.
The former has a major impact on observing efficiency,
as a large number of spectra can be fit on each CCD frame if the
wavelength coverage is restricted. For the clusters considered here, this was
done using a filter which only passed light from 4300 to 5600\AA,
corresponding roughly to 3500 to 4550\AA\ in the z=0.2 rest frame.
Moderately deep $g$ and $R$ images were also recorded of each field
to identify objects for slit mask construction. Additional details of
the observations will be given by Yee et al. (in preparation).

	The clusters discussed in the present paper lie at the low redshift
end of the CNOC sample, which was selected such that the x-ray luminosity
between 1 and 3 Kev was at least $5 \times 10^{44}$ erg sec$^{-1}$.
The most luminous cluster galaxies at z=0.2 are
moderately bright and well resolved, practical considerations which are
essential for studies of spatial photometric and spectroscopic properties.
The six z=0.2 clusters for which spectra of the BCG were obtained are Abell
2390, EMSS0440+02, 0839+29, 0906+11, 1231+15,
and 1455+22. The BCG's were identified from the $g$
and $R$ images recorded of each field, as well as from the atlas of
images compiled by Goia \& Luppino (1994). Aside from the BCG's, a small
number of other bright cluster galaxies, selected based on their having clearly
extended spectra, were also examined. An observing log is given in Table 1.

	The images were reduced using standard techniques, which included bias
subtraction and division by a dome flat. Portions of the final $g$
images containing the BCG and surrounding galaxies are shown in Figures
$1 - 6$. The galaxies with spectra discussed in this paper are labelled, with
$A$ refering to the BCG, $B$ to the second brightest galaxy, etc. It is
evident from these figures that the BCG's are located in a range of
environments. EMSS0440a, EMSS0839a, and EMSS0906a have one or more bright
sub-nuclei within $\sim 5$ arcsec. The remaining systems do not show clearly
resolved sub-nuclei, although the outer isophotes of
Abell2390a are extended towards the North West.
A faint tidal tail appears to connect EMSS1231a and b, which have
a projected separation of $\sim 17$ arcsec.

	The spectroscopic data were reduced using procedures similar to those
described in earlier studies of nearby galaxies (e.g. Davidge 1992; Davidge \&
Clark 1994), the only difference being that each of the present exposures
contained multiple spectra. A median bias image was subtracted from each
CCD frame, which typically contained spectra of 60 or more objects.
Two-dimensional spectra of individual galaxies
were extracted from the results. These were
divided by flat-field frames, created by directing light from a
continuum lamp through the slit masks.
The flat-field frames provided corrections for pixel-to-pixel
variations in detector sensitivity, as well as for vignetting
along the slit. The flat-fielded spectra were then wavelength calibrated
using He-Ar arc and bright sky emission lines. The sky level was measured from
nearby slits containing only faint galaxies.

	One-dimensional spectra were extracted by co-adding along the spatial
axis, with the data being folded about the photometric center of each galaxy to
increase the signal-to-noise ratio in the faint outer regions. The typical
bin size was $\sim 0.9$ arcsec, although wider intervals were
used for the fainter systems. The extracted spectra were pseudo-flux calibrated
using observations of the spectrophotometric flux standard HZ 44 (Oke 1990).
Rest frame spectra of individual galaxies, summed along the slit,
are compared in Figures 7, 8, 9, and 10.

\begin{center}
3. MEASUREMENT OF SPECTRAL INDICES
\end{center}

	Velocities for each galaxy were computed from the centroids
of Ca H \& K, the G-band and, when detected, [OII]
$\lambda 3727$. The results are listed in the second column of Table 2.
There is a large velocity difference between EMSS0440a and EMSS0440b and c,
a result not uncommon to BCG's and nearby companions (e.g. Tonry
1985). In contrast, EMSS0906a and c have very similar velocities.

	The primary spectral index adopted for the investigation of stellar
content in the present study is D4000, which measures the strength of the
4000\AA\ break based on the ratio of fluxes in the wavelength intervals $3750 -
3950$\AA\ and $4050 - 4250$\AA\ (Bruzual 1983).
As in Davidge \& Clark (1994), D4000 is given in magnitude units,
rather than as a flux ratio. D4000 is a useful probe of stellar content
since it is sensitive to both age and metallicity (van den Bergh 1963:
van den Bergh \& Sackmann 1965, but see also Dressler \&
Shectman 1987). Moreover, it is computed from relatively broad
passbands, and can be measured in spectra with relatively low S/N ratios,
making it an ideal diagnostic for the faint outer regions of galaxies.

	To provide supplemental spectral information, we also calculated
CN4170 (Burstein et al. 1984), an index which measures the depth of CN
absorption near 4200\AA. CN4170 shows considerable scatter when compared with
other indices in composite stellar systems, such as
Mg$_2$ (Burstein et al. 1984); nevertheless, like D4000 it has
the advantage of being computed from moderately broad passbands. Another useful
feature is that the continuum points bracket the CN bands, so that the index is
insensitive to reddening, which is not the case for D4000.

	The spectra of Abell2390a, EMSS0440b, EMSS0839a, and EMSS1455a
contain [OII]$\lambda 3727$ emission. The strength of this
feature was measured using the index defined by Dressler \&
Shectman (1987), with the result given as an equivalent width, in \AA.

	Galaxy-wide D4000 and CN4170 indices, computed from the spatially
integrated spectra shown in Figures 7, 8, 9, and 10, are listed in Table 2,
while measurements made within various distance intervals are listed in Table
3. The [OII] indices for EMSS0839, EMSS1455, and Abell2390a are
listed in Table 4. The two-dimensional spectrum of EMSS0440b reveals a
compact emission line region East of the galaxy center, and [OII]
measurements are also given for this region in Table 4.

	Uncertainties in the D4000 and CN4170 indices were estimated from the
column-to-column scatter in the EMSS0440a data. The uncertainty in D4000 is
$\sim 0.07$ magnitudes near the galaxy center, and $\sim 0.15$ magnitudes in
the outermost regions. In the case of CN4170, the uncertainties range from
$\sim 0.02$ magnitudes near the center to $\sim 0.08$ magnitudes in the outer
regions. These uncertainties also apply to the other systems in our sample.

	If the CN4170 and D4000 indices listed in Table 3 monitor a common
parameter, such as age or metallicity, then they should be correlated. To
investigate if this is the case, CN4170 was plotted as a function of D4000, and
the result is shown in the upper panel of Figure 11. Indices from the outermost
bins were not included, as these have the largest uncertainties. The
CN4170 and D4000 indices for EMSS0839a, EMSS1455a, and Abell2390a are also not
shown, given that the spectra of these systems contain strong
[OII] emission. It is evident that the data in Figure 11 define a common trend,
in the sense that CN4170 increases with D4000. It is interesting that the data
from galaxies in EMSS0440 and EMSS1231 are in good agreement, even though
these systems show systematically different color gradients (Section 5).

	The complex relation between D4000 and CN4170 defined by the present
observations is similar to that in nearby galaxies.
This is demonstrated in the lower panel of Figure 11, where we compare data
from Table 3 with observations for nearby galaxies and
metal rich solar neighborhood stars from Davidge \& Clark (1994).
Those portions of nearby galaxies containing measureable
[OII] emission have been excluded from this comparison, as have data from the
outermost bins. It is evident that the nearby and
z=0.2 galaxy datasets overlap and define similar trends when D4000
$\geq 0.6$. It is also apparent that the measurements for the intermediate
redshift galaxies extend to lower values of D4000 and CN4170, although the
significance of this finding is not clear, as the spectra obtained by Davidge
\& Clark (1994) did not sample the outermost regions of nearby
galaxies. Nevertheless, in the region not covered by nearby galaxy
data, the trend between CN4170 and D4000 defined by the z=0.2
galaxies parallels that defined by solar neighborhood stars,
indicating that the measured indices are physically reasonable.

	The comparison in the lower panel of Figure 11 suggests that common
parameters control the stellar contents of nearby and intermediate
redshift galaxies. The stellar contents of the z=0.2 galaxies can be
further investigated by comparing the central D4000 indices from
Table 3 with those in corresponding regions of nearby elliptical galaxies. A
comparison of this nature requires that aperture effects be taken into account,
since a fixed aperture samples progressively larger projected areas as redshift
increases, with the consequence that the mean metallicity within this aperture
drops with increasing redshift. Davidge \& Clark (1994) used long-slit spectra
of two nearby elliptical galaxies, NGC2693 and NGC4486 (M87), to model the
appearance of systems at intermediate redshift, as viewed through an aperture
with 1 arcsec radius during 1 arcsec seeing conditions. A D4000 index
corresponding to this aperture size was derived for each galaxy in the current
sample by averaging the D4000 indices in the two centermost bins. The results
are compared with the predictions made by Davidge \& Clark (1994) in Figure 12.
It is apparent that the central D4000 indices for the EMSS0440 and EMSS1231
systems are similar to those expected for NGC2693, which is in many respects a
`typical' elliptical galaxy (Davidge \& Clark 1994). However, the central D4000
indices for EMSS0839a, EMSS1455a, and Abell2390a are substantially lower than
those predicted for NGC2693, a result that is not surprising given the
widespread [OII] emission in these systems, and the accompanying blue
continuum. The comparison in Figure 12 indicates that those galaxies in our
sample without line emission have global stellar contents similar to
nearby elliptical galaxies; however, in Section 5 it will be shown that the
central regions of the EMSS0440 galaxies contain a blue component not seen in
the other galaxies.

\begin{center}
4. COLOR MEASUREMENTS
\end{center}

	A large database of $g-r$ measurements exists for
low and intermediate redshift BCG's (Schneider et al. 1983a).
Therefore, prior to computing colors, the instrumental $R$ brightnesses were
transformed into Thuan \& Gunn (1976) $r$ values. Background levels
were measured from the central portions of each image using the $SKY$
routine in DAOPHOT (Stetson 1987).

	While photometric data on its own provides some insight into stellar
content, a detailed comparison between broad-band colors and spectroscopic
indices is of interest as the latter typically provide information on the
strengths of individual atomic and/or molecular species,
whereas the former give information on global system properties, such as the
color of the red giant branch, brightness of the main-sequence turn-off, and
extinction. A comparison of this nature requires a
direct spatial correspondence between the
photometric and spectroscopic datasets. In particular,
rather than measure colors over large regions of each galaxy, the analysis
should be restricted only to those areas covered by the spectrograph slit.
This can be readily achieved with the present observations, because the
spectroscopic and photometric data were recorded with the same instrument and
detector. The integrated $g-r$ colors within the area covered by the slit
for each galaxy are listed in Table 2, while individual color measurements as
functions of distance from the galaxy centers are listed in Table 3. The
estimated uncertainties in the color measurements are modest;
in the case of EMSS0440a, the uncertainties are $\sim 0.01$ magnitudes near the
galaxy center, and $\sim 0.02$ magnitudes in the outermost bin.

	Galaxy-to-galaxy color comparisons are complicated by differences in
redshift. To compensate for this, the k-corrections computed by Schneider et
al. (1983a) were used to calculate rest frame $g-r$ colors, which we denote by
$(g-r)_0$, and the results are listed in Table 3. The Schneider et al.
k-corrections were derived from nearby elliptical galaxies using an
instrumental configuration which had poor blue response, and an effective $g$
wavelength much redder than the standard value. Although it is difficult to
reproduce the observing system used by Schneider et al. (1983a), there
are indications that these k-corrections are valid for our
dataset. In particular, the comparisons in Figure 11 and 12
suggest that the stellar contents of the EMSS0440, EMSS0906, and EMSS1231
galaxies are similar to those of nearby ellipticals. Hence, we would expect the
k-corrected colors of these galaxies to be similar to those of nearby galaxies.
Schneider et al. (1983a) adopted a template spectral-energy distribution (SED)
for nearby galaxies which had $g-r \sim 0.47$, and the k-corrected colors in
Tables 2 and 3 are very similar to this value.
Nevertheless, we caution that the corrected colors are
sensitive to stellar content. Moreover, we note that the elliptical galaxy
k-corrections predicted by Frei \& Gunn (1994) would produce colors $\sim 0.06$
magnitudes redder than those computed here. K-corrections
were not applied to EMSS0869a, EMSS1455a, and
Abell2390a, as it is likely that these galaxies have SED's very different
from the others. Nevertheless, it is apparent from Tables 2 and 3 that $g-r$ in
the BCG's showing [OII] emission tends to be bluer than the other systems with
similar redshifts.

	D4000 and CN4170 are plotted as functions of $(g-r)_0$ in Figure 13,
where the measurements for the outermost bins have been excluded to reduce
scatter, as have the EMSS0839a, EMSS1455a, and Abell2390a observations. It is
evident that the spectral indices and $(g-r)_0$ are not related in a simple
fashion, and that the data for each cluster may define a separate trend. Part
of the scatter in the horizontal direction may be due to uncertainties in the
k-corrections. For example, the k-corrections computed by Frei \& Gunn (1994)
suggest that $\sim 0.1$ magnitude smearing in $(g-r)_0$ would be introduced if
some of the galaxies had stellar contents similar to those of early-type
spirals rather than elliptical galaxies. Uncertainties
in stellar content notwithstanding, the data plotted
in Figure 13 ostensibly suggest that spectroscopic and broad-band
photometric measurements may be sensitive to different parameters, a finding
supported by the lack of correlation between color and spectroscopic gradients
(Section 5). Clearly, a combination of both spectroscopic and photometric
information is desireable to draw conclusions about stellar content, even in
systems which appear to have stellar contents similar to those of nearby
ellipticals.

\begin{center}
5. EVIDENCE FOR POPULATION GRADIENTS
\end{center}

\noindent{\it 5.1 Characterization of Gradients}

	In Figures 14 $-$ 19 we show D4000, CN4170, and $(g-r)_0$ as
functions of log(r), where r is the distance from the galaxy center in arcsec,
for all 11 galaxies in our sample. Inspection of these Figures reveals a number
of interesting trends. First, the behaviours of D4000 and $g-r$ with radius in
the BCG's exhibiting [OII] emission are very different from what is seen in the
other BCG's. Second, among the systems not showing [OII] emission there is a
clear tendency for the spectroscopic features to weaken with increasing radius,
and when log(r) $\geq 0.0$ (ie. outside the seeing disk) monotonic trends with
log(r) are evident. Finally, the radial behaviour of $(g-r)_0$ exhibits
systematic cluster-to-cluster differences. In particular, all three
galaxies in EMSS0440 show blue central regions, and a tendency to become
redder with increasing radius until log(r) $\sim 0.0 - 0.5$, at which point the
colors become progressively bluer. For comparison, a central blue population is
not seen in the EMSS0906 and EMSS1231 galaxies. We note that these color
profile differences could not be predicted from the spectroscopic measurements
alone, and this further demonstrates the need for combined spectroscopic and
photometric datasets to investigate stellar content.

	In an effort to characterize any trends,
least squares linear fits were made to the spectroscopic indices
and $(g-r)$ colors as functions of log(r), and the resulting values of
$\Delta$D4000/$\Delta$log(r), $\Delta$CN4170/$\Delta$log(r), and
$\Delta$$(g-r)$/$\Delta$log(r) are listed in Table 5. Indices measured within
the seeing disk were excluded from the analysis, as these
points would cause the computed gradients to be underestimated. Least squares
fits could not be applied to EMSS0839a and EMSS0906c because of the
limited spatial coverage of these systems. Moreover, we
emphasize that although linear fits do not provide an ideal
representation of the radial behaviour of $(g-r)_0$ in the EMSS0440
galaxies, they nevertheless provide a means of quantifying
deviations from nearby galaxies, where linear color
gradients are seen (e.g. Peletier et al. 1990).

\vspace{0.5cm}
\noindent{\it 5.2 Comparison with spectroscopic gradients in nearby systems
and other BCG's}

	Over the past decade, spectroscopic gradients have been studied in a
moderately large number of nearby elliptical galaxies, and the resulting
measurements provide a potentially important source of reference
for the present work. Unfortunately, many of these studies concentrated
on the rest frame optical wavelength region (e.g. Couture \&
Hardy 1988; Gorgas, Efstathiou, \& Aragon Salamanca 1990; Davidge 1992;
Davies, Sadler, \& Peletier 1993; Carollo, Danziger, \& Buson 1993), so there
is little or no wavelength overlap with the present dataset.
However, Davidge \& Clark (1994) investigated the near-ultraviolet
and blue wavelength regions of six nearby elliptical galaxies, and found that
D4000 gradients are common, with $\Delta$D4000/$\Delta$log(r) ranging from
$-0.090$ to $-0.484$. Davidge \& Clark (1994) also found a real spread in
$\Delta$D4000/$\Delta$log(r), in the sense that the galaxy-to-galaxy dispersion
is larger than predicted by measurement errors.
A similar result has also been found for $\Delta$Mg$_2$/$\Delta$log(r)
(Gorgas et al. 1990; Davidge 1992). Finally, few of the nearby
systems studied by Davidge \& Clark 1994 have significant CN4170 gradients.

	There are some similarities between the spectroscopic gradients
seen in nearby galaxies and those measured in the current intermediate-redshift
sample. Significant (which we define as having slopes different from zero in
excess of the $3\sigma$ level) D4000 gradients are seen in EMSS0440a and b,
EMSS0906a and b, and EMSS1231a. The $\Delta$D4000/$\Delta$log(r) values
measured for the EMSS0440, 0906, and 1231 galaxies fall within the range
seen by Davidge \& Clark (1994), although there is
a tendency to favour the high end of the nearby galaxy range.
Furthermore, it is evident from Table 5 that the CN4170 gradients are either
poorly defined or flat, and only EMSS1231a has $\Delta$CN4170/$\Delta$log(r)
significantly different from zero. Finally, a possible difference between the
current sample and nearby systems concerns the behaviour of line indices at
large radii. Gorgas et al. (1990) and Davies et al. (1993) find
a tendency for line gradients in some (but not all) nearby
systems to flatten at large radii. None of the galaxies in our sample
show this behaviour.

	Many of the nearby galaxies which have measured spectroscopic
gradients are relatively faint (ie. M$_B \sim -20$), and are not located in
dense cluster environments. Therefore, it is important to consider how the
radial spectroscopic properties of the current sample compare with what is seen
among nearby bright cluster galaxies. The closest and most extensively studied
galaxy showing cD-like characteristics is M87, for which Davidge \& Clark
(1994) and Davidge (1992) measured relatively steep D4000 and Mg$_2$ gradients
when compared with other nearby ellipticals. Davies et al. (1993) measured a
significantly shallower Mg$_2$ gradient in M87, although they
still assign this galaxy one of the steepest Mg$_2$ gradients in their
elliptical sample. The properties of M87 notwithstanding, it is likely that
this galaxy is not representative of all bright cluster ellipticals.
For example, based on a sample of three cD's, Gorgas et al. (1990)
found $\Delta$Mg$_2$/$\Delta$log(r) values similar to nearby
elliptical galaxies. Moreover, McNamara \& O'Connell (1989) measured the
strength of the 4200\AA\ CN bands in a sample of cooling-flow BCG's spanning a
range of redshifts. They also conclude that the metallicity gradients in these
systems are similar to those in nearby ellipticals, although the
failure to detect significant CN4170 gradients in the present sample and the
systems studied by Davidge \& Clark (1994) suggests that the blue CN bands may
not provide an ideal means of investigating population gradients.

	The uncertainties in the $\Delta$D4000/$\Delta$log(r)
measurements in Table 5 are relatively large. Therefore, in an effort to search
for systematic trends, the EMSS0440, 0906, and 1231 galaxies were sorted into
two groups: (1) BCG's only (ie. EMSS0440a, 0906a, and 1231a), and
(2) non-BCG's.  Values of $\Delta$D4000/$\Delta$log(r) for each of
these categories were computed after shifting the measurements
along the spatial axis to compensate for differences in scale length.
The results, summarized in Table 6, suggest that the D4000 gradients
among the BCG's may be slightly steeper than among the non-BCG's, although the
significance of this result is low.

\vspace{0.5cm}
\noindent{\it 5.3 Comparison with color gradients in nearby systems}

	Based solely on the numbers in Table 5, only EMSS1231a and EMSS0440c
show significant color gradients, with $g-r$ becoming bluer with increasing
radius. However, it is clear from Figures 18 and 19 that all three EMSS0440
galaxies show systematic radial color variations. Modest color
gradients are relatively common in nearby elliptical galaxies (e.g.
Schombert et al. 1993; Peletier et al. 1990). A study of particular relevance
is that by Cohen (1986), who investigated $g-r$ gradients in three bright
Virgo cluster galaxies, and found that $\Delta (g-r)$/$\Delta$log(r)
$\sim -0.04$. It is unfortunate that this slope is comparable to the
uncertainties given in the last column of Table 5.

	There are indications that the color gradients in BCG's and cD's may be
different from those in nearby ellipticals, although caution should be
exercised in comparisons of this nature given the results in Figure 18, which
indicate that galaxies with [OII] emission
can have very different colors gradients than those with no emission.
Schombert (1988), Mackie (1992) and Andreon et al. (1992) used CCD data to
study color gradients in cD's and BCG's out to intermediate redshifts, and
found that the optical colors typically do not
vary with radius. McNamara \& O'Connell (1992) studied the colors of 19
centrally dominant cooling flow galaxies, and found that in many cases the
color profiles are either flat or positive (ie. redder colors occuring at
larger radii). The presence of a significant linear color gradient in
EMSS1231a, and the hint of a drop off in $(g-r)_0$ when log(r) $\geq 0.5$ in
the EMSS0440 galaxies are ostensibly inconsistent with these generalized
results; however, we note that Mackie (1992) found a large $g-r$ gradient in
the Abell2634 D galaxy NGC7728. Based on the data in his Figure 3, we
estimate that $\Delta (g-r)_0$/$\Delta$log(r) $\sim -0.2$, in rough agreement
with what is seen in EMSS1231a and the outer regions of EMSS0440c.

\begin{center}
6. DISCUSSION
\end{center}

\noindent{\it 6.1 Comments on Individual Systems}

	The EMSS0440 and EMSS0906 systems are of interest because the BCG's
have bright companions located at small projected distances from the main
galaxy. Tonry (1985) investigated the dynamics of BCG companions, and
concluded that most of these must lie on highly elliptical orbits to satisfy
the observed spatial distributions and velocity dispersions. If
this is the case, then BCG companions should have high velocities
when they pass close to the larger galaxy, and it might then be anticipated
that any tidal interactions will be slight, so that these galaxies would emerge
relatively unscathed from encounters with the BCG (Tonry 1985). Some support
for this picture comes from the CCD images discussed by Porter et al. (1991),
which reveal that the structural properties of BCG's are independent of whether
or not close companions are present.

	Is there evidence for recent interactions among the EMSS0440 and EMSS\-0906
galaxies? While the spectroscopic and photometric
properties of the EMSS\-0906 galaxies do not appear to be peculiar,
there are some indications that the EMSS0440
galaxies might have experienced recent interactions. In particular,
a patch of bright [OII] emission, presumably due to star formation
activity, is seen to one side of EMSS0440b. The
redshift of this material is identical to that of EMSS0440b,
and we speculate that this region may ultimately evolve into a low surface
brightness object similar to those detected in the outer envelopes of BCG's by
Porter et al. (1991). Moreover, all three EMSS0440 galaxies show a blue
nucleus, although the spectra of the central regions of the these galaxies do
not show strong Balmer absorption features.

	Inspection of highly stretched images of EMSS1231a and b suggest that
these galaxies are connected by a faint tidal bridge. Despite this visual
evidence for interaction, the integrated colors and spectral indices of
EMSS1231a and b are not peculiar, and there are no indications of strong
Balmer absorption. Consequently, it appears that recent large-scale star
formation has not occured in either galaxy. One possible peculiar aspect of
EMSS1231a is that it contains a very steep $g-r$ color gradient, although we
emphasize that this system is not unique, as Mackie (1992) has found another
galaxy with similar properties.

	Half of the BCG's in our sample show strong [OII] emission,
an ostensibly surprising result since the fraction of nearby BCG's
showing strong [OII] emission is low. For example,
Schneider et al. (1983a) found only one BCG out of their total sample of 84
which exhibited an emission spectrum. The high incidence of [OII] emission
is probably a consequence of the x-ray brightness criterion used to select
the sample. Indeed, Johnstone, Fabian, \& Nulsen (1987) studied a sample of
galaxies located at the centers of cooling flows, and found a much higher
incidence of [OII] emission. Based on the strengths of various spectroscopic
features, Johnstone et al. conclude that star formation provides the most
reasonable source of excitation in these systems.

	EMSS0839a, EMSS1455a, and Abell2390a have properties
which suggest that they may be systematically different
from the galaxies studied by Johnstone et al. In particular, the mean
D4000 index in the Johnstone et al. (1987) sample containing measureable [OII]
emission is $\sim 0.71$ magnitudes, and this is much higher than what is
seen in the z=0.2 emission line BCG's studied here. Furthermore, Johnstone
et al. (1987) measured central and global D4000 indices, so their data
can be used to place loose constraints on D4000 gradients in cooling flow
galaxies. On average, their data indicate that D4000 does not change
with radius when [OII] emission is present, whereas
there is a definite tendency for D4000 to {\it strengthen} with
increasing distance from the center of Abell2390a, and {\it weaken} with
increasing distance from the center of EMSS0839a.

\vspace{0.5cm}
\noindent{\it 6.2 D4000 and Surface Brightness}

	Photometric and spectroscopic surveys of elliptical galaxies have
revealed that mean metallicity, as measured either from color (e.g. Sandage \&
Visvanathan 1978) or line strength (e.g. Terlevich et al. 1981),
and galaxy brightness are related. Models of galaxy formation suggest that
this trend has its origin in the mass-dependence of galactic escape velocities
(e.g. Dekel \& Silk 1986), in the sense that more massive systems are able to
retain gas longer in the presence of supernovae-driven winds,
permitting their interstellar mediums, and hence stellar populations, to become
more chemically evolved. It has recently been demonstrated that a similar trend
holds within individual galaxies, where local color (Franx \& Illingworth 1990)
and line strengths (Davies et al. 1993) are related to local escape velocity.

	Given these indications that the chemical enrichment histories
of regions within galaxies depend on escape velocity,
which in turn depends on local mass density, gradients measured in terms of
surface brightness, rather than distance from the galaxy center, may provide a
more fundamental means of comparing the evolution of elliptical galaxies.
Therefore, local surface brightnesses for the z=0.2 galaxies were calculated
from the $g$ images at various points along the slits,
and the results are listed in the third column of
Table 3. The relation between D4000 and SB$_g$ for each galaxy is
shown in Figures 20 and 21, and monotonic trends between these quantities are
clearly evident. Least squares linear fits were applied to determine
$\Delta$D4000/$\Delta$SB$_g$ for each system, and the results are summarized in
Table 5. Significant gradients are present in 6 galaxies; moreover, there
does not appear to be significant galaxy-to-galaxy differences in
$\Delta$D4000/$\Delta$SB$_g$ among the systems
free of [OII] emission. This latter point is demonstrated further in
Figure 22, which shows D4000 and SB$_g$ for the galaxies in EMSS0440, 0906, and
1231. The scatter among D4000 values at a fixed surface brightness is roughly
consistent with that predicted from measurement errors (Section 4).
Nevertheless, the EMSS0906 data largely defines the lower envelope of the
distribution. The slope computed from the data in Figure 22 is
$\Delta$D4000/$\Delta$SB$_g \sim -0.12 \pm 0.02$. The fact that the data in
Figure 22 falls within a well-defined region suggests that the various
galaxies in our sample experienced similar chemical enrichment histories.

	Baum, Thomsen, \& Morgan (1986) and Thomsen \& Baum (1987, 1989) used
narrow band images to measure $\Delta$Mg$_2$/$\Delta$SB$_V$ in three Coma
cluster ellipticals, including the cD NGC4874. They found only modest
galaxy-to-galaxy differences in the gradients
measured in this manner, and found a mean value of
$\Delta$Mg$_2$/$\Delta$SB$_V \sim -0.04$. In order to compare this value with
that derived for the galaxies in EMSS0440, 0906, and 1231, we estimated
$\Delta$Mg$_2$/$\Delta$D4000 and $\Delta$SB$_g$/$\Delta$SB$_V$ from nearby
elliptical galaxy observations. The former was computed by combining the
$\Delta$D4000/$\Delta$log(r) values measured by Davidge \& Clark (1994) for 6
nearby elliptical galaxies with $\Delta$Mg$_2$/$\Delta$log(r) values measured
for the same systems by Davidge (1992). Individual $\Delta$Mg$_2$/$\Delta$D4000
values for these six calibrating galaxies are listed in Table 7. It is
interesting to note that $\Delta$Mg$_2$/$\Delta$D4000 for M87 is not
significantly different from the other systems, even though the stellar content
of this system at large radii appears to be different from that of other
elliptical galaxies (e.g. Davidge \& Clark 1994; McNamara \& O'Connell 1989).
The mean $\Delta$Mg$_2$/$\Delta$D4000 value from Table 7 is $0.344 \pm 0.047$.
As for estimating $\Delta$SB$_V$/$\Delta$SB$_g$, the transformation equations
given by Kent (1985) indicate that $\Delta$V $\sim \Delta$g at fixed $B-V$, so
that $\Delta$SB$_V$/$\Delta$SB$_g \sim 1$. Consequently, multiplying
$\Delta$Mg$_2$/$\Delta$SB$_V$ by $\Delta$D4000/$\Delta$Mg$_2$ and
$\Delta$SB$_V$/$\Delta$SB$_g$, we find that $\Delta$D4000/$\Delta$SB$_g \sim
-0.12 \pm 0.02$ for bright Coma cluster ellipticals, in excellent agreement
with that derived for EMSS0440, 0906, and 1231 galaxies.

\vspace{0.5cm}
\noindent{\it 6.3 The Relation between D4000 and Metallicity}

	Dressler \& Shectman (1987) investigated the behaviour of the 4000\AA\
break among low redshift cluster galaxies, and found only a weak trend with
galaxy brightness, leading them to conclude that the
4000\AA\ break is relatively insensitive to metallicity, at
least in the super-metal rich regime. The detection of D4000 gradients in
nearby elliptical galaxies by Davidge \& Clark (1994), coupled with the result
that $\Delta$D4000/$\Delta$Mg$_2 \sim 2.9 \pm 0.4$ (Section 6.2), suggests that
the metallicity dependence of D4000 may be stronger than originally
suggested by Dressler \& Shectman (1987). We have estimated the metallicity
sensitivity of D4000 in order to compute $\Delta$[Fe/H]/$\Delta$log(r), an
important statistic for comparison with galaxy formation models. A fundamental
assumption in the procedure described below is that the stellar contents of the
EMSS0440, 0906, and 1231 galaxies outside of their central regions are similar
to those of nearby ellipticals, and the comparisons in Sections 3, 4, and 5
suggests that this assumption is reasonable.

	The metallicity sensitivity of D4000 can be estimated by combining the
$\Delta$Mg$_2$/$\Delta$D4000 value calculated in Section 6.1 with
theoretical relations linking Mg$_2$ and [Fe/H]. For the present work,
we consider the Mg$_2 -$ [Fe/H] calibrations given by Worthey, Faber, \&
Gonzalez (1992) and Buzzoni, Gariboldi, \& Mantegazza (1992). There is
considerable disagreement between these calibrations; for example, near solar
metallicity, the former conclude that $\Delta$Mg$_2$/$\Delta$[Fe/H]
= 0.26, whereas the latter find 0.135. The metallicity sensitivities predicted
by these calibrations, found by multiplying
$\Delta$D4000/ $\Delta$Mg$_2$ by $\Delta$Mg$_2$/$\Delta$[Fe/H], are
$\Delta$D4000/$\Delta$[Fe/H] = $0.76 \pm 0.10$ (Worthey et al.) and
$0.39 \pm 0.05$ (Buzzoni et al.), where the quoted uncertainties reflect
only the errors in the relation between D4000 and Mg$_2$. It should be
emphasized that these results assume linear relations between Mg$_2$ and
[Fe/H], an assumption which likely breaks down at low metallicities.

	The metallicity gradients derived from these calibrations are
given in the last two columns of Table 6.
Metallicity gradients are usually interpreted as a
signature of dissipation, and their presence implies that gas was present at
some point during the collapse of either the individual galaxies or, if they
formed in a hierarchal fashion, their progenitors. The
metallicity gradients in Table 6 are steeper than those in most nearby
elliptical galaxies (e.g. Davidge 1992), suggesting that the galaxies in
EMSS0440, 0906, and 1231 experienced comparatively large amounts of
dissipation.

\begin{center}
7. SUMMARY
\end{center}

	Spectroscopic and broad-band photometric observations, obtained with
the same instrument and sampling the same spatial regions, have been used to
investigate the radial properties of eleven bright galaxies in the z=0.2
clusters EMSS0440+02, 0839+29, 0906+11, 1231+15, 1455+22, and Abell2390. We
find that (1) the stellar contents of the BCGs in EMSS0839, EMSS1455, and
Abell2390a are very different from those of the others in our sample,
in that they show strong [OII] emission and a relatively blue continuum;
(2) the integrated photometric and spectroscopic properties of the
remaining systems suggest that their global stellar contents are similar to
those of nearby elliptical galaxies, although the galaxies in EMSS0440
contain a centrally-concentrated blue component; (3)
there is a general tendency for the 4000\AA\ break to weaken with
increasing distance from the galaxy centers in those systems which do not
show [OII] emission, indicating that stellar content
changes with radius; (4) the measured values of $\Delta$D4000/$\Delta$log(r)
fall within the range seen in nearby elliptical galaxies; (5) there is a slight
tenedency for the steepest gradients to occur in the BCG's; (6) radial
variations in $g-r$ occur in most systems, although the trends seen in the
EMSS0440 galaxies are very different from those in the EMSS0906 and EMSS1231
galaxies; (6) there is not a clear correlation between broad-band color and
line strength, suggesting that these quantities are sensitive to different
parameters; (7) the galaxies in EMSS0440, EMSS0906, and
EMSS1231 define a common locus in the D4000 $-$ SB$_g$ plane, which argues for
similar chemical enrichment histories; and (8) using data from Davidge \&
Clark (1994), we find that D4000 is sensitive to [Fe/H], contrary to the
conclusions reached by Dressler \& Shectman (1987).

	The manner with which stellar populations change with radius
provides a fossil record of the collapse histories of galaxies. White (1978;
1980) modelled the merging of equal mass stellar systems, and found
that the end products have population gradients which are
flatter than those of their progenitors; for example, after three
merging events, White finds that any gradients are flattened by a
factor of 2. Hence, the presence of relatively steep line gradients in the
galaxies studied here indicates that they likely did not form from the
merger of equal mass gas poor systems. A model in which a large number of gas
poor low mass systems are accreted by a larger seed galaxy also seems unlikely,
as the end product should also contain relatively flat metallicity gradients,
due to the absence of dissipation following the accretion events. We speculate
that the homogeneity seen on the D4000 $-$ SB$_g$ plane may also be hard to
create with mergers of chemically evolved systems of varying masses
(and possibly mass-to-light ratios), although
this will require detailed simulations to establish quantitatively.

	The simulations discussed above did not include the effects of gas and
star formation. However, it is not clear if
significant numbers of gas-rich systems are present
in the central regions of evolved rich clusters given, for example, the
existence of a morphology-density relation (e.g. Dressler 1980). Nevertheless,
gas-rich systems were very likely present in the centers of protoclusters
at high redshift $-$ otherwise, galaxies would not have formed. Therefore,
the detection of steep spectroscopic gradients
lead us to {\it speculate} that the main bodies of EMSS0440a,
0906a, and 1231a formed at early epochs. Caution should be exercised when
attempting to generalize these results to other BCG's, since the CNOC clusters
were selected based on their x-ray luminosities. It would be of interest to
determine if BCG's in x-ray poor clusters have radial properties different
from those in x-ray bright systems.

	We close by noting that the infered metallicity gradients are
{\it roughly} consistent with those predicted by theoretical simulations.
Carlberg (1984) found that dissipative monolithic collapse models could produce
systems with metallicity gradients as steep as $\Delta$[Fe/H]/$\Delta$log(r)
$\sim -0.6$, in rough agreement with the metallicity gradient predicted for the
BCG's in EMSS0440, 0906 and 1231 using the Worthey et al. (1992) metallicity
calibration. We also note that, based on the data presented here, EMSS0440a,
EMSS0839a, EMSS\-1455a, and Abell2390a likely contain relatively large young
populations. Studies of BCG's at greater redshifts will be of interest to see
if the fraction of systems with similar properties increases with lookback
time.

\vspace{1.0cm}
	It is a pleasure to thank the CNOC observing team: Howard Yee, Erica
Ellingson, Ray Carlberg, Tammy Smecker-Hane, Bob Abraham, Mike Rigler, Simon
Morris, and Chris Pritchet for their dedicated efforts.
Sincere thanks are extended to Sidney van den Bergh, Simon
Morris, and Ray Carlberg for commenting on an earlier version of the
manuscript. An anonymous referee also made suggestions that greatly
improved this paper. TJD gratefully acknowledges support from the
Natural Sciences Engineering Research Council of Canada (NSERC) and the
National Research Council of Canada (NRC).

\pagebreak[4]
\begin{center}
TABLE 1. Observing Log
\end{center}

\begin{center}
\begin{tabular}{llll}
\hline\hline
Cluster & Date & Filter & t$_{exp}$ \\
 & Observed & & (sec) \\
\hline
Abell2390 & June 18, 1993 & $g$ & 900 \\
 & June 18, 1993 & $R$ & 900 \\
 & June 22, 1993 & Spectra & 4080 \\
 & & & \\
EMSS0440+02 & January 14, 1994 & $g$ & 600 \\
 & January 14, 1994 & $R$ & 600 \\
 & January 16, 1994 & Spectra & 3600 \\
 & & & \\
EMSS0839+29 & March 8, 1994& $g$ & 600 \\
 & March 8, 1994& $R$ & 600 \\
 & March 14, 1994& Spectra & 3300 \\
 & & & \\
EMSS0906+11 & March 8, 1994 & $g$ & 600 \\
 & March 8, 1994 & $R$ & 600 \\
 & March 12, 1994 & Spectra & 3600 \\
 & & & \\
EMSS1231+15 & January 14, 1994 & $g$ & 600 \\
 & January 14, 1994 & $R$ & 600 \\
 & January 15, 1994 & Spectra & 3600 \\
 & & & \\
EMSS1455+22 & March 11, 1994 & $g$ & 600 \\
 & March 11, 1994 & $R$ & 600 \\
 & March 13, 1994 & Spectra & 3600 \\
\hline\hline
\end{tabular}
\end{center}

\pagebreak[4]
\begin{center}
TABLE 2. Galaxy Properties
\end{center}

\begin{center}
\begin{tabular}{lcrrrr}
\hline\hline
Galaxy & $cz$ & D4000 & CN4170 & $g-r$ & $(g-r)_0$\\
 & (km/sec) & & & & \\
\hline
EMSS0440a & 60000 & 0.695 & 0.115 & 0.83 & 0.46 \\
EMSS0440b & 56000 & 0.751 & 0.067 & 0.78 & 0.43 \\
EMSS0440c & 56000 & 0.714 & 0.130 & 0.71 & 0.36 \\
EMSS0839a & 57900 & 0.384 & 0.138 & 0.77 & $-$ \\
EMSS0906a & 52800 & 0.596 & 0.077 & 0.76 & 0.45 \\
EMSS0906b & 52500 & 0.628 & 0.008 & 0.76 & 0.45 \\
EMSS0906c & 52200 & 0.676 & 0.030 & 0.74 & 0.43 \\
EMSS1231a & 70000 & 0.674 & 0.020 & 0.95 & 0.45 \\
EMSS1231b & 70000 & 0.751 & 0.063 & 0.95 & 0.45 \\
EMSS1455a & 76800 & 0.648 & 0.023 & 0.75 & $-$ \\
Abell2390a & 69000 & 0.515 & 0.064 & 0.72 & $-$ \\
\hline\hline
\end{tabular}
\end{center}

\pagebreak[4]
\begin{center}
TABLE 3. Absorption Line Indices and Broad-Band Colors
\end{center}

\begin{center}
\begin{tabular}{llllrll}
\hline\hline
Galaxy & Radii & SB$_g$ & D4000 & CN4170 & $g-r$ & $(g-r)_0^a$ \\
 & (arcsec) &  & (mag) & (mag) & & \\
\hline
EMSS0440a & $0.0 - 0.5 $ & 21.139 & 0.857 & 0.194 & 0.75 & 0.38 \\
 & $0.5 - 1.4$ & 21.555 & 0.853 & 0.113 & 0.87 & 0.50 \\
 & $1.4 - 2.3$ & 22.135 & 0.599 & 0.167 & 0.88 & 0.51 \\
 & $2.3 - 3.2$ & 22.516 & 0.750 & 0.144 & 0.90 & 0.53 \\
 & $3.2 - 4.2$ & 22.909 & 0.572 & $-0.067$ & 0.90 & 0.53 \\
 & $4.2 - 5.1$ & 23.334 & 0.414 & 0.100 & 0.87 & 0.50 \\
 & $5.1 - 6.0$ & 23.711 & 0.325 & $-0.049$ & 0.85 & 0.48 \\
 & & & & & \\
EMSS0440b & $0.0 - 0.5$ & 21.237 & 0.872 & 0.106 & 0.71 & 0.36 \\
 & $0.5 - 1.4$ & 21.626 & 0.769 & 0.111 & 0.81 & 0.46 \\
 & $1.4 - 2.3$ & 22.123 & 0.690 & 0.056 & 0.85 & 0.50 \\
 & $2.3 - 3.2$ & 22.992 & 0.591 & $-0.196$ & 0.86 & 0.51 \\
 & $3.2 - 5.1$ & 23.664 & 0.613 & $-0.046$ & 0.82 & 0.47 \\
 & & & & & \\
EMSS0440c & $0.0 - 0.5 $ & 21.684 & 0.813 & 0.124 & 0.64 & 0.29 \\
 & $0.5 - 1.4$ & 22.306 & 0.719 & 0.194 & 0.81 & 0.46 \\
 & $1.4 - 2.3$ & 23.094 & 0.689 & 0.030 & 0.77 & 0.42 \\
 & $2.3 - 3.2$ & 23.899 & 0.350 & $-0.019$ & 0.74 & 0.39 \\
 & & & & & \\
EMSS0839a & $0.0 - 0.5 $ & 21.639 & 0.455 & 0.176 & 0.81 & $-$ \\
 & $0.5 - 1.4$ & 22.180 & 0.361 & 0.113 & 0.75 & $-$ \\
 & $1.4 - 2.3$ & 22.861 & 0.334 & 0.153 & 0.75 & $-$ \\
\hline\hline
\end{tabular}
\end{center}

\vspace{1.0cm}
\noindent{$^a$} rest frame $g-r$ colors, as computed using the k-corrections
derived by Schneider et al. (1983a).

\pagebreak[4]
\begin{center}
TABLE 3. (con't)
\end{center}

\begin{center}
\begin{tabular}{llllrll}
\hline\hline
Galaxy & Radii & SB$_g$ & D4000 & CN4170 & $g-r$ & $(g-r)_0^a$ \\
 & (arcsec) & & (mag) & (mag) & & \\
\hline
EMSS0906a & $0.0 - 0.5$ & 21.337 & 0.771 & 0.101 & 0.78 & 0.47 \\
 & $0.5 - 1.4$ & 21.807 & 0.654 & 0.072 & 0.77 & 0.46 \\
 & $1.4 - 2.3$ & 22.327 & 0.548 & 0.073 & 0.73 & 0.42 \\
 & $2.3 - 3.2$ & 22.396 & 0.454 & 0.064 & 0.77 & 0.46 \\
 & $3.2 - 4.2$ & 22.884 & 0.411 & 0.070 & 0.73 & 0.42 \\
 & & & & & \\
EMSS0906b & $0.0 - 0.5$ & 20.741 & 0.727 & 0.031 & 0.79 & 0.48 \\
 & $0.5 - 1.4$ & 21.513 & 0.689 & 0.023 & 0.75 & 0.44 \\
 & $1.4 - 2.3$ & 22.488 & 0.437 & $-0.118$ & 0.74 & 0.43 \\
 & $2.3 - 3.2$ & 23.222 & 0.412 & $-0.033$ & 0.69 & 0.38 \\
 & $3.2 - 4.2$ & 22.786 & 0.335 & 0.148 & 0.72 & 0.41 \\
 & & & & & \\
EMSS0906c & $0.0 - 0.5$ & 21.432 & 0.806 & $-0.006$ & 0.77 & 0.46 \\
 & $0.5 - 1.4$ & 22.051 & 0.639 & 0.072 & 0.72 & 0.41 \\
 & $1.4 - 2.9$ & 23.328 & 0.484 & 0.009 & 0.73 & 0.42 \\
 & & & & & \\
EMSS1231a & $0.0 - 0.5$ & 22.027 & 0.758 & 0.055 & 1.08 & 0.58 \\
 & $0.5 - 1.4$ & 22.341 & 0.791 & 0.074 & 0.96 & 0.46 \\
 & $1.4 - 2.3$ & 22.969 & 0.577 & $-0.029$ & 0.79 & 0.29 \\
 & $2.3 - 3.2$ & 23.667 & 0.427 & $-0.095$ & 0.75 & 0.25 \\
 & $3.2 - 4.2$ & 24.177 & 0.502 & $-0.078$ & 0.72 & 0.22 \\
\hline\hline
\end{tabular}
\end{center}

\vspace{1.0cm}
\noindent{$^a$} rest frame $g-r$ colors, as computed using the k-corrections
derived by Schneider et al. (1983a).

\pagebreak[4]
\begin{center}
TABLE 3. (con't)
\end{center}

\begin{center}
\begin{tabular}{llllrll}
\hline\hline
Galaxy & Radii & SB$_g$ & D4000 & CN4170 & $g-r$ & $(g-r)_0^a$ \\
 & (arcsec) & & (mag) & (mag) & & \\
\hline
 & & & & & \\
EMSS1231b & $0.0 - 0.5$ & 22.140 & 0.884 & 0.159 & 1.04 & 0.54 \\
 & $0.5 - 1.4$ & 22.450 & 0.715 & 0.102 & 0.94 & 0.44 \\
 & $1.4 - 2.3$ & 23.182 & 0.797 & 0.109 & 0.85 & 0.35 \\
 & $2.3 - 3.2$ & 23.899 & 0.368 & $-0.447$ & 0.72 & 0.22 \\
 & $3.2 - 5.1$ & 24.781 & 0.690 & $-0.360$ & 0.89 & 0.39 \\
 & & & & & \\
EMSS1455a & $0.0 - 0.5$ & 21.468 & 0.687 & 0.033 & 0.77 & $-$ \\
 & $0.5 - 1.4$ & 21.966 & 0.631 & 0.017 & 0.75 & $-$ \\
 & $1.4 - 2.3$ & 22.855 & 0.611 & 0.016 & 0.79 & $-$ \\
 & $2.3 - 3.9$ & 23.717 & 0.626 & 0.013 & 0.76 & $-$ \\
 & $3.9 - 5.7$ & 23.992 & 0.658 & 0.037 & 0.66 & $-$ \\
 & & & & & \\
Abell2390a & $0.0 - 0.5$ & 21.206 & 0.481 & 0.060 & 0.73 & $-$ \\
 & $0.5 - 1.4$ & 21.733 & 0.517 & 0.110 & 0.69 & $-$ \\
 & $1.4 - 2.3$ & 22.728 & 0.484 & 0.030 & 0.70 & $-$ \\
 & $2.3 - 3.4$ & 23.468 & 0.667 & 0.042 & 0.72 & $-$ \\
 & $3.4 - 4.2$ & 23.922 & 0.559 & $-0.228$ & 0.81 & $-$ \\
\hline\hline
\end{tabular}
\end{center}

\vspace{1.0cm}
\noindent{$^a$} rest frame $g-r$ colors, as computed using the k-corrections
derived by Schneider et al. (1983a).

\pagebreak[4]
\begin{center}
TABLE 4. [OII] Equivalent Widths
\end{center}

\begin{center}
\begin{tabular}{llr}
\hline\hline
Galaxy & Radii & [OII] \\
 & (arcsec) & (\AA) \\
\hline
EMSS0440b & $0.5 - 1.4$ (East) & 12.2 \\
 & $1.4 - 2.3$ (East) & 41.2 \\
 & & \\
EMSS0839a & $0.0 - 0.5$ & 51.2 \\
 & $0.5 - 1.4$ & 32.7 \\
 & $1.4 - 2.3$ & 8.3 \\
 & & \\
EMSS1455a & $0.0 - 0.5$ & 27.2 \\
 & $0.5 - 1.4$ & 24.4 \\
 & $1.4 - 2.3$ & 10.7 \\
 & $2.3 - 3.9$ & 4.0 \\
 & $3.9 - 5.7$ & 19.5 \\
 & & \\
Abell2390a & $0.0 - 0.5$ & 61.4 \\
 & $0.5 - 1.4$ & 38.4 \\
 & $1.4 - 2.3$ & 18.3 \\
 & $2.3 - 3.4$ & 11.0 \\
 & $3.4 - 4.2$ & 3.8 \\
\hline\hline
\end{tabular}
\end{center}

\pagebreak[4]
\begin{center}
TABLE 5. Computed Gradients
\end{center}

\begin{center}
\begin{tabular}{lcccc}
\hline\hline
Galaxy & $\frac{\Delta D4000}{\Delta log(r)}$ & $\frac{\Delta CN4170}{\Delta
log(r)}$ & $\frac{\Delta g-r}{\Delta log(r)}$ & $\frac{\Delta D4000}{\Delta
SB_g}$ \\
\hline
EMSS0440a & $-0.60 \pm 0.17$ & $-0.20 \pm 0.14$ & $-0.02 \pm 0.02$ & $-0.21 \pm
0.04$ \\
EMSS0440b & $-0.27 \pm 0.07$ & $-0.34 \pm 0.24$ & $+0.02 \pm 0.06$ & $-0.11 \pm
0.03$ \\
EMSS0440c & $-0.70 \pm 0.48$ & $-0.45 \pm 0.07$ & $-0.15 \pm 0.01$ & $-0.19 \pm
0.06$ \\
EMSS0906a & $-0.41 \pm 0.03$ & $-0.01 \pm 0.01$ & $-0.03 \pm 0.05$ & $-0.24 \pm
0.03$ \\
EMSS0906b & $-0.57 \pm 0.10$ & $+0.15 \pm 0.28$ & $-0.09 \pm 0.05$ & $-0.14 \pm
0.02$ \\
EMSS1231a & $-0.55 \pm 0.17$ & $-0.28 \pm 0.06$ & $-0.39 \pm 0.07$ & $-0.16 \pm
0.04$ \\
EMSS1231b & $-0.23 \pm 0.45$ & $-0.87 \pm 0.43$ & $-0.15 \pm 0.21$ & $-0.10 \pm
0.09$ \\
EMSS1455a & $+0.04 \pm 0.04$ & $+0.02 \pm 0.02$ & $-0.11 \pm 0.10$ & $-0.10 \pm
0.04$ \\
Abell2390a & $+0.16 \pm 0.18$ & $-0.45 \pm 0.24$ & $+0.16 \pm 0.10$ & $+0.04
\pm 0.03$ \\
\hline\hline
\end{tabular}
\end{center}

\pagebreak[4]
\begin{center}
TABLE 6. Mean Gradients
\end{center}

\begin{center}
\begin{tabular}{llll}
\hline\hline
Sample & ${\Delta D4000/\Delta log(r)}$ & ${\Delta [Fe/H]/\Delta log(r)}^a$ &
${\Delta [Fe/H]/\Delta log(r)}^b$ \\
\hline
Bcg's & $-0.54 \pm 0.08$ & $-0.7 \pm 0.1$ & $-1.4 \pm 0.3$ \\
Non-Bcg's & $-0.44 \pm 0.11$ & $-0.6 \pm 0.2$ & $-1.1 \pm 0.3$ \\
\hline\hline
\end{tabular}
\end{center}

\vspace{1.0cm}
\noindent{$^a$} Worthey et al. (1992) calibration

\noindent{$^b$} Buzzoni et al. (1992) calibration

\pagebreak[4]
\begin{center}
TABLE 7. $\Delta$Mg$_2$/$\Delta$D4000 Values
\end{center}

\begin{center}
\begin{tabular}{ll}
\hline\hline
Galaxy & $\Delta$Mg$_2$/$\Delta$D4000 \\
\hline
NGC2693 & $0.256 \pm 0.108$ \\
NGC3348 & $0.411 \pm 0.215$ \\
NGC3379 & $0.533 \pm 0.148$ \\
NGC4278 & $0.344 \pm 0.111$ \\
NGC4486 & $0.304 \pm 0.023$ \\
NGC5044 & $0.214 \pm 0.084$ \\
\hline\hline
\end{tabular}
\end{center}

\pagebreak[4]
\parindent=0.0cm
\begin{center}
REFERENCES
\end{center}

Andreon, S., Garilli, B., Maccagni, D., Gregorini, L., \& Vettolani, G.
\linebreak[4]\hspace*{2.0cm}1992, A\&A, 266, 127

Baum, W. A., Thomsen, B., \& Morgan, B. L. 1986, ApJ, 301, 83

Bhavsar, S. P. 1989, ApJ, 338, 718

Bingelli, B. 1982, A\&A, 107, 338

Blakeslee, J. P., \& Tonry, J. L. 1992, AJ, 103, 1457

Bruzual, A. G. 1983, ApJS, 42, 565

Burstein, D., Faber, S. M., Gaskell, C. M., \& Krumm, N. 1984, ApJ, 287,
\linebreak[4]\hspace*{2.0cm}586

Buzzoni, A., Gariboldi, G., \& Mantegazza, L. 1992, AJ, 103, 1814

Carlberg, R. G. 1984, ApJ, 286, 403

Carlberg, R. G. et al. 1994, JRASC, 88, 39

Carollo, C. M., Danziger, I. J., \& Buson, L. 1993, MNRAS, 265, 553

Carter, D., \& Metcalfe, N. 1981, MNRAS, 191, 325

Cohen, J. G. 1986, AJ, 92, 1039

Couture, J., \& Hardy, E. 1988, AJ, 96, 867

Davidge, T. J. 1992, AJ, 103, 1512

Davidge, T. J., \& Clark, C. C. 1994, AJ, 107, 946

Davies, R. L., Sadler, E. M., \& Peletier, R. F. 1993, MNRAS, 262, 650

Dekel, A., \& Silk, J. 1986, ApJ, 303, 39

Dressler, A. 1980, ApJ, 236, 351

Dressler, A., \& Shectman, S. A. 1987, AJ, 94, 899

Franx, M., \& Illingworth, G. 1990, ApJ, 359, L41

Goia, I., \& Luppino, J. 1994, ApJS, 94, 583

Gorgas, J., Efstathiou, G., \& Aragon Salamanca, A. 1990, MNRAS, 245, 217

Hausman, M. A., \& Ostriker, J. P. 1978, ApJ, 224, 320

Hoessel, J. G., Gunn, J. E., \& Thuan, T. X. 1980, ApJ, 241, 486

Johnstone, R. M., Fabian, A. C., \& Nulsen, P. E. J. 1987, MNRAS, 224, 75

Katz, N., \& White S. D. M. 1993, ApJ, 412, 455

Lauer, T. R., \& Postman, M. 1992, ApJL, 400, L47

Lee, M. G., \& Geisler, D. 1993, AJ, 106, 493

Le F\`{e}vre, O., Crampton, D., Felenbok, P., \& Monnet, G. 1993, A\&A, 282,
\linebreak[4]\hspace*{2.0cm}325

Mackie, G. 1992, ApJ, 400, 65

McLaughlin, D. E., Harris, W. E., \& Hanes, D. A. 1993, ApJL, 409, L45

McNamara, B. R., \& O'Connell, R. W. 1989, AJ, 98, 2018

McNamara, B. R., \& O'Connell, R. W. 1992, ApJ, 393, 579

Merritt, D. 1985, ApJ, 289, 18

Morbey, C. L. 1992, Applied Optics, 31, 2291

Oegerle, W. R., \& Hoessel, J. G. 1991, ApJ, 375, 15

Oke, J. B. 1990, AJ, 99, 1621

Peletier, R., Davies, R., Illingworth, G., Davies, L., \& Cawson, M. 1990,
\linebreak[4]\hspace*{2.0cm}AJ, 100, 1091

Porter, A. C., Schneider, D. P., \& Hoessel, J. G. 1991, AJ, 101, 1561

Rhee, G., van Haarlem, M., \& Katgert, P. 1992, AJ, 103, 1721

Ryden, B. S., Lauer, T. R., \& Postman, M. 1993, ApJ, 410, 515

Sandage, A., \& Hardy, E. 1973, ApJ, 183, 743

Sandage, A., \& Visvanathan, N. 1978, ApJ, 223, 707

Sastry, G. N. 1968, PASP, 80, 252

Schneider, D. P., Gunn, J. E., \& Hoessel, J. G. 1983a, ApJ, 264, 337

Schneider, D. P., Gunn, J. E., \& Hoessel, J. G. 1983b, ApJ, 268, 476

Schombert, J. M. 1988, ApJ, 328, 475

Schombert, J. M., Hanlan, P. C., Barsony, M., \& Rakos, K. D. 1993, AJ, 106,
\linebreak[4]\hspace*{2.0cm}923

Stetson, P. B. 1987, PASP, 99, 191

Struble, M. F. 1987, ApJ, 317, 668

Terlevich, R., Davies, R. L., Faber, S. M., \& Burstein, D. 1981, MNRAS, 196,
\linebreak[4]\hspace*{2.0cm}381

Thomsen, B., \& Baum, W. A. 1987, ApJ, 315, 460

Thomsen, B., \& Baum, W. A. 1989, ApJ, 347, 214

Thuan, T. X., \& Gunn, J. E. 1976, PASP, 88, 543

Tonry, J. L. 1985, ApJ, 291, 45

van den Bergh, S. 1963, AJ, 68, 413

van den Bergh, S., \& Sackmann, I. J. 1965, AJ, 70, 353

White, S. D. M. 1978, MNRAS, 184, 185

White, S. D. M. 1980, MNRAS, 191, 1P

Worthey, G., Faber, S. M., \& Gonzalez, J. J. 1992, ApJ, 398, 69

\pagebreak[4]
\begin{center}
FIGURE CAPTIONS
\end{center}

\parindent=0.0cm
Figure 1: $g$ image of the central 2 x 2 arcmin of the cluster EMSS0440+02.
The three galaxies studied in this paper are labelled, with $A$ being the BCG.
North is at the top and West is to the right.

\vspace{0.5cm}
Figure 2: Same as Figure 1, but for EMSS0839+29.

\vspace{0.5cm}
Figure 3: Same as Figure 1, but for EMSS0906+11.

\vspace{0.5cm}
Figure 4: Same as Figure 1, but for EMSS1231+15.

\vspace{0.5cm}
Figure 5: Same as Figure 1, but for EMSS1455+22.

\vspace{0.5cm}
Figure 6: Same as Figure 1, but for Abell2390.

\vspace{0.5cm}
Figure 7: Spatially integrated spectra of the BCG's not showing [OII] emission.
The wavelength scale is in the rest frame. Note that the spectra have been
normalised and offset vertically for the purposes of display.

\vspace{0.5cm}
Figure 8: Same as Figure 7, but for BCG's showing [OII] emission.

\vspace{0.5cm}
Figure 9: Spatially integrated spectra of the galaxies which are close
companions to BCG's. Note the weak [OII] $\lambda 3727$ emission in EMSS0440b,
which originates from a concentrated region to the East of the galaxy center.
The wavelength scale is in the rest frame. The spectra have been normalised
and offset vertically for the purposes of display.

\vspace{0.5cm}
Figure 10: Same as Figure 9, but for cluster galaxies located at
moderately large distances from the BCG's.

\vspace{0.5cm}
Figure 11: Relation between CN4170 and D4000 as defined by z=0.2 cluster
galaxies, nearby ellipticals, and metal-rich solar neighborhood stars. The
upper panel shows data for EMSS0440 (crosses), EMSS0906 (open triangles), and
EMSS1231 (solid squares) galaxies. These z=0.2 galaxy data
are plotted as crosses in the lower panel, while nearby elliptical
galaxy observations are shown as open squares. Indices for metal-rich
solar neighborhood dwarfs (filled triangles) and giants (open triangles) are
also shown in the lower panel. Note that the trends defined by nearby
elliptical galaxies and solar neighborhood stars are consistent with that
defined by z=0.2 galaxies.

\vspace{0.5cm}
Figure 12: Relationship between D4000 and red shift predicted for NGC2693
(solid line) and NGC4486 (M87 - dashed line) if these galaxies were shifted
to larger distances and viewed through 1 arcsec
radius apertures during 1 arcsec seeing conditions (Davidge \& Clark
1994). Also shown are the central D4000 values for
EMSS0440 (cross), EMSS0839 (open triangle), EMSS0906 (filled triangle),
EMSS1231 (open square), EMSS1455 (filled square), and Abell2390 (star)
galaxies.

\vspace{0.5cm}
Figure 13: Relationships between k-corrected $g-r$ color, $(g-r)_0$, and
spectral indices for EMSS0440 (crosses), EMSS0906 (open triangles) and EMSS1231
(filled squares) galaxies. The top and bottom panels show D4000 and
CN4170, respectively.

\vspace{0.5cm}
Figure 14: D4000 as a function of log(r) for the six BCG's. The top row
contains those systems which do not show [OII] emission and these are, from
left to right, EMSS0440a, EMSS0906a, and EMSS1231a. The bottom row contains
those BCG's which exhibit [OII] emission and these are, from left to right,
EMSS0839a, EMSS1455a, and Abell2390a. Errorbars are shown for the first and
last point for EMSS0440a. The dashed lines are the least squares linear fits
obtained by excluding the central point.

\vspace{0.5cm}
Figure 15: D4000 as a function of log(r) for galaxies which are not BCG's. The
top row contains those systems which are close companions to BCG's which are,
from left to right, EMSS0440b, EMSS0440c, and EMSS0906c. The bottom row
contains those galaxies which are located at moderately large distances from
BCG's which are EMSS0906b (left) and EMSS1231b (right).
The dashed lines are the least squares linear fits
obtained by excluding the central point.

\vspace{0.5cm}
Figure 16: CN4170 as a function of log(r) for the six BCG's. The top row
contains those systems which do not show [OII] emission and these are, from
left to right, EMSS0440a, EMSS0906a, and EMSS1231a. The bottom row contains
those BCG's which exhibit strong [OII] emission which are, from left to right,
EMSS0839a, EMSS1455a, and Abell2390a. Errorbars are shown for the first and
last point for EMSS0440a. The dashed lines are the least squares linear fits
obtained by excluding the central point.

\vspace{0.5cm}
Figure 17: CN4170 as a function of log(r) for galaxies which are not BCG's. The
top row contains those systems which are close companions to BCG's which are,
from left to right, EMSS0440b, EMSS0440c, and EMSS0906c. The bottom row
contains those galaxies which are located at moderately large distances from
BCG's which are EMSS0906b (left) and EMSS1231b (right).
The dashed lines are the least squares linear fits
obtained by excluding the central point.

\vspace{0.5cm}
Figure 18: $(g-r)_0$ as a function of log(r) for the six BCG's. The top row
contains those systems which do not show [OII] emission and these are, from
left to right, EMSS0440a, EMSS0906a, and EMSS1231a. The bottom row contains
those BCG's which exhibit strong [OII] emission which are, from left to right,
EMSS0839a, EMSS1455a, and Abell2390a.

\vspace{0.5cm}
Figure 19: $(g-r)_0$ as a function of log(r) for galaxies which are not BCG's.
The top row contains those systems which are close companions to BCG's which
are, from left to right, EMSS0440b, EMSS0440c, and EMSS0906c. The bottom row
contains those galaxies which are located at moderately large distances from
BCG's which are EMSS0906b (left) and EMSS1231b (right).

\vspace{0.5cm}
Figure 20: SB$_g$ as a function of log(r) for the six BCG's. The top row
contains those systems which do not show [OII] emission and these are, from
left to right, EMSS0440a, EMSS0906a, and EMSS1231a. The bottom row contains
those BCG's which exhibit strong [OII] emission which are, from left to right,
EMSS0839a, EMSS1455a, and Abell2390a. Errorbars are shown for the first and
last point for EMSS0440a. The dashed lines are the least squares linear fits
obtained by excluding the central point.

\vspace{0.5cm}
Figure 21: SB$_g$ as a function of log(r) for galaxies which are not BCG's.
The top row contains those systems which are close companions to BCG's which
are, from left to right, EMSS0440b, EMSS0440c, and EMSS0906c. The bottom row
contains those galaxies which are located at moderately large distances from
BCG's which are EMSS0906b (left) and EMSS1231b (right).
The dashed lines are the least squares linear fits
obtained by excluding the central point.

\vspace{0.5cm}
Figure 22: Relation between D4000 and SB$_g$ for EMSS0440 (crosses),
EMSS\-0906 (open triangles), and EMSS1231 (filled squares) galaxies.
\end{document}